\documentclass[11pt,aps,prd,showpacs,floatfix,nofootinbib,superscriptaddress]{revtex4-2}

\usepackage{graphicx}  % needed for figures
\usepackage{multirow}
\usepackage[dvipsnames]{xcolor}
\usepackage[colorlinks=true,allcolors=BlueViolet]{hyperref}
\usepackage{latexsym}
\usepackage{amsfonts,amsmath,amssymb}
\usepackage{slashed}
\usepackage{color}
\usepackage{orcidlink}
\usepackage{supertabular} 
\usepackage{epsfig}
\usepackage{lipsum}

\newcommand*{\nr}{\text{nr}}

\newcommand*{\Tr}{\text{Tr}}
\newcommand*{\bs}{\boldsymbol}

\newcommand{\olsi}[1]{\,\overline{\!{#1}}}

\newcommand{\itp}{\affiliation{Institute of Theoretical Physics, Chinese Academy of Sciences, Beijing 100190, China}}

\newcommand{\ucas}{\affiliation{School of Physical Sciences, University of Chinese Academy of Sciences, Beijing 100049, China}}

\newcommand{\ific}{\affiliation{Instituto de F\'isica Corpuscular (centro mixto CSIC-UV), \\
Institutos de Investigaci\'on de Paterna, Apartado 22085, 46071, Valencia, Spain}}

\newcommand{\peng}{\affiliation{Peng Huanwu Collaborative Center for Research and Education,\\ Beihang University, Beijing 100191, China}}
\newcommand{\scnt}{\affiliation{Southern Center for Nuclear-Science Theory (SCNT), Institute of Modern Physics,\\ Chinese Academy of Sciences, Huizhou 516000, China}}

\usepackage{tikz}
\usepackage{epstopdf}
\usetikzlibrary{trees}
\usetikzlibrary{arrows.meta,shapes,graphs}
\usetikzlibrary{matrix,arrows} 			
\usetikzlibrary{positioning}				
\usetikzlibrary{calc,through}				
\usetikzlibrary{decorations.pathreplacing}  
\usepackage{pgffor}							
\usetikzlibrary{decorations.pathmorphing}	
\usetikzlibrary{decorations.markings}
\tikzset{
    vector/.style={decorate, decoration={snake}, draw},
	provector/.style={decorate, decoration={snake,amplitude=2.5pt}, draw},
	antivector/.style={decorate, decoration={snake,amplitude=-2.5pt}, draw},
    fermion/.style={draw=black, postaction={decorate},
        decoration={markings,mark=at position .55 with {\arrow[draw=black]{>}}}},
    fermionbar/.style={draw=black, postaction={decorate},
        decoration={markings,mark=at position .55 with {\arrow[draw=black]{<}}}},
    fermionnoarrow/.style={draw=black},
    gluon/.style={decorate, draw=black,
        decoration={coil,amplitude=4pt, segment length=5pt}},
    scalar/.style={dashed,draw=black, postaction={decorate},
        decoration={markings,mark=at position .55 with {\arrow[draw=black]{>}}}},
    scalarbar/.style={dashed,draw=black, postaction={decorate},
        decoration={markings,mark=at position .55 with {\arrow[draw=black]{<}}}},
    scalarnoarrow/.style={dashed,draw=black},
    electron/.style={draw=black, postaction={decorate},
        decoration={markings,mark=at position .55 with {\arrow[draw=black]{>}}}},
    bigvector/.style={decorate, decoration={snake,amplitude=4pt}, draw},
    bound/.style={decorate,decoration={zigzag},draw},
}
\tikzstyle{block} = [draw, rectangle, 
    minimum height=3em, minimum width=6em]
\begin{document}

\title{Production of $1^{-+}$ exotic charmonium-like states in electron-positron collisions}

\author{Xiao-Yu~Zhang}\email{zhangxiaoyu@itp.ac.cn} \itp \ucas

\author{Pan-Pan~Shi\orcidlink{0000-0003-2057-9884}}\email{Panpan.Shi@ific.uv.es}
\ific
\author{Feng-Kun~Guo\orcidlink{0000-0002-2919-2064}}\email{fkguo@itp.ac.cn}
\itp\ucas\peng\scnt

\begin{abstract}
The absence of observed charmonium-like states with the exotic quantum numbers $J^{PC}=1^{-+}$ has prompted us to investigate the production rates of the $1^{-+}$ $D\olsi D_1(2420)$ and $D^*\olsi D_1(2420)$ hadronic molecules, which we refer to as $\eta_{c1}$ and $\eta_{c1}^{\prime}$, respectively, in electron-positron collisions. Assuming a hadronic molecular nature for the vector charmonium-like states $\psi(4360)$ and $\psi(4415)$, we evaluate the radiative decay widths of $\psi(4360)\to\gamma\eta_{c1}$ and $\psi(4415)\to\gamma\eta_{c1}^{\prime}$. 
Using these decay widths, we estimate the cross sections for producing $\eta_{c1}$ and $\eta_{c1}^{\prime}$ in electron-positron annihilations, as well as the event numbers at the planned Super $\tau$-Charm Facility. Our results suggest that the ideal energy region for observing these states is around $4.44$ and $4.50$~GeV, just above the $D^* \olsi D_1(2420)$ and $D^*\olsi D_2^*(2460)$ thresholds, respectively. 
\end{abstract}

\maketitle

\section{Introduction}

In the last two decades, high-energy experiments have reported a plethora of hadronic structures. Many of these structures are not consistent with the conventional quark-antiquark mesons and three-quark baryons, and are considered as candidates for exotic states. Numerous experimental and theoretical studies have been conducted on these exotic states, as reviewed in recent literature~\cite{Hosaka:2016pey,Esposito:2016noz,Lebed:2016hpi,Ali:2017jda,Olsen:2017bmm,Guo:2017jvc,Liu:2019zoy,Guo:2019twa,Brambilla:2019esw,Chen:2022asf,ParticleDataGroup:2024cfk,Liu:2024uxn,Chen:2024eaq,Wang:2025sic}. 
Among these, states with exotic quantum numbers $J^{PC}$, such as $J^{PC}=0^{--}$, $0^{+-}$, $1^{-+}$, $2^{+-}$, are particularly intriguing. The absence of the quark-antiquark component in these states simplifies the study of their underlying structure. 

In the light-quark sector, several exotic states with quantum numbers $J^{PC}=1^{-+}$ have been observed in high-energy experiments. The isovector states $\pi_{1}(1400)$ and $\pi_1(1600)$ were reported with the exotic quantum numbers $J^{PC}=1^{-+}$~\cite{ParticleDataGroup:2024cfk,Meyer:2015eta}. 
Furthermore, the BESIII Collaboration discovered the isoscalar state $\eta_1(1855)$ with $J^{PC}=1^{-+}$ through the decay process $J/\psi \to \gamma\eta_1(1855)\to \gamma \eta\eta^{\prime}$~\cite{BESIII:2022riz,BESIII:2022iwi}. Nevertheless, to date, no evidence of an exotic $J^{PC}$ state has been confirmed in the hidden-charm sector.

In the context of lattice quantum chromodynamics (QCD), the hidden-charm exotic states with quantum numbers $J^{PC}=1^{-+}$ have been predicted as a hybrid in both quenched~\cite{Liao:2002rj,Mei:2002ip,Dudek:2008sz,Ma:2019hsm} and unquenched calculations~\cite{HadronSpectrum:2012gic,Shi:2023sdy}. 
In contrast, Ref.~\cite{Yang:2012gz} presents a calculation of the vacuum-to-meson matrix elements for both the $1^{-+}$ meson and ordinary mesons, using the quenched approximation and dimension-4 $\olsi \psi \gamma_4 \overset{\leftrightarrow{}}{D}_i \psi$ and dimension-5 $\olsi \psi \varepsilon_{ijk}\gamma^j\psi B^k$ interpolating operators. 
The comparable contributions of these matrix elements to exotic and ordinary mesons suggest that there is no evidence supporting the hybrid nature of the $1^{-+}$ meson~\cite{Yang:2012gz}. Additionally, the unquenched calculation~\cite{HadronSpectrum:2012gic} reveals that the $1^{-+}$ meson lies approximately $72(16)$ MeV below the $D\olsi D_1(2420)$ threshold, even without incorporating charmed-meson interpolating operators.\footnote{This calculation employed a pion mass of about 400 MeV. 
Here, we disregard the light quark mass dependence of this exotic state and utilize the physical isospin-average masses of $D$, $\olsi D_1$, and $\eta_c$ to estimate the binding energy, based on the mass splitting between the $1^{-+}$ meson and $\eta_{c}$ reported in Ref.~\cite{HadronSpectrum:2012gic}. } 
Notably, the binding energy of this $1^{-+}$ meson is comparable to that of $\psi(4230)$.\footnote{Assuming the $\psi(4230)$ to be a $D\olsi D_1(2420)$ bound state~\cite{Ding:2008gr,Li:2013bca,Wang:2013cya}, its binding energy is about 64~MeV~\cite{ParticleDataGroup:2024cfk}.} 
The calculation in Ref.~\cite{Shi:2023sdy} shows that the $1^{-+}$ exotic meson has a strong coupling to $D_1\bar D$ though the predicted mass is slightly above the $D_1\bar D$ threshold.
These findings indicate that the $1^{-+}$ meson is a strong candidate for a $D\olsi D_1$ hadronic molecule.

Extensive studies have been conducted on the $1^{-+}$ states within the framework of hadronic molecules. Predictions for $1^{-+}$ molecular-like resonances were made based on the Coulomb gauge model~\cite{General:2007bk}. The mass of the $1^{-+}$ $D\olsi D_1(2420)$ bound state was estimated using the chiral SU(3) quark model~\cite{Li:2013bca}. 
Subsequently, its binding energy and partial widths into various channels were estimated using vector-meson-exchange potentials~\cite{Dong:2019ofp}. The mass spectrum of three possible $1^{-+}$ molecular states, associated with the $D\olsi D_1(2420)$, $D^*\olsi D_1(2420)$, and $D^*\olsi D_2^*(2460)$ thresholds~\cite{Cleven:2015era}, was predicted using the light vector-meson-exchange model~\cite{Dong:2021juy}.

The lack of a $1^{-+}$ meson signal in the hidden-charm sector underscores the need for theoretical studies to support experimental searches in high-energy experiments. 
Electron-positron collisions offer a promising avenue for searching for the exotic $1^{-+}$ mesons. The BESIII Collaboration has measured the cross sections for the process $e^+e^-\to \gamma X(3872)$ at center-of-mass (c.m.) energies $\sqrt s = 4.009$, $4.229$, $4.260$, and $4.360$~GeV~\cite{BESIII:2013fnz}, confirming theoretical prediction made in the molecular picture in Ref.~\cite{Guo:2013zbw}, but no signal was detected in the energy range of 4.66 to 4.95~GeV~\cite{BESIII:2024ext}. 
Furthermore, in the process $e^+e^-\to\gamma X$, no $C$-even charmonium-like states $X$ were observed in the $D_s^{\pm}D_s^{*\mp}$ invariant mass distribution at $\sqrt s =(4681.92\pm 0.30)$ MeV~\cite{BESIII:2024tbj}. These BESIII measurements imply that a search for the $1^{-+}$ mesons in $e^+e^-$ collisions is most feasible at $\sqrt{s} < 4.66$ GeV. 
In addition, the Super $\tau$-Charm Facility (STCF) under discussion is expected to have an luminosity two orders of magnitude higher than the BESIII experiment~\cite{Achasov:2023gey}, which would make it more promising for searching for the $1^{-+}$ exotic states in $e^+e^-$ collisions.
Under the assumption that the ground-state $1^{-+}$ meson is a $D\olsi D_1(2420)$ hadronic molecule, denoted as $\eta_{c1}$ in this paper, its production in the reaction $e^+e^-\to \psi(4360)\to \gamma \eta_{c1}$ has been estimated in Ref.~\cite{Wang:2014wga}. The predicted cross section for this process is approximately 0.1~pb at $\sqrt s=4.36$~GeV. 
To further explore the production of $1^{-+}$ exotic mesons within the hadronic molecular framework, encompassing the $D\olsi D_1(2420)$ and $D^* \olsi D_1(2420)$ molecules, with the latter denoted as $\eta_{c1}^{\prime}$, we calculate the decay widths $\psi(4360)\to \gamma \eta_{c1}$ and $\psi(4415)\to \gamma\eta_{c1}^{\prime}$. In our analysis, we regard $\psi(4360)$ and $\psi(4415)$ as $D^*\olsi D_1(2420)$ and $D^*\olsi D_2^*(2460)$ hadronic molecules, respectively~\cite{Wang:2013kra,Ma:2014zva,Cleven:2015era,Hanhart:2019isz,Anwar:2021dmg}. 
In this case, the production could be sizeable because of the large effective coupling of a hadronic molecule to its constituents.
Based on the calculated radiative decay widths, we employ the charmonium-like vector meson dominance model (VMD) to estimate the cross sections for the processes $e^+e^-\to\psi(4360)\to\gamma\eta_{c1}$ and $e^+e^-\to\psi(4415)\to\gamma\eta_{c1}^{\prime}$.

This paper is organized as follows. In Sec.~\ref{Sec:formalism}, we present the Lagrangian and amplitudes for the radiative decays $\psi(4360)\to\gamma\eta_{c1}$ and $\psi(4415)\to\gamma\eta_{c1}^{\prime}$. The radiative decay widths and angular distributions are discussed in Sec.~\ref{Sec:decay_width}. 
In Sec.~\ref{Sec:production}, we estimate the cross sections for the production of $\eta_{c1}$ and $\eta_{c1}^{\prime}$ in $e^+e^-$ annihilations. A conclusion is given in Sec.~\ref{Sec:conlusion}. 
In Appendix~\ref{Sec:polarization}, we list the polarization vectors in the nonrelativistic approximation. Finally, Appendix~\ref{Sec:decay_X3872} contains an estimate of the production rate for the process $e^+e^-\to\psi(4360)\to\gamma X(3872)$.

\section{Formalism}\label{Sec:formalism}

\subsection{Effective Lagrangian}

In the hadronic molecular picture, the $\psi(4360)$ and $\psi(4415)$ are considered to be the $ D_1\olsi D^*$ and $D^*\olsi D_2^*$ molecules, respectively, with the quantum numbers $J^{PC}=1^{--}$~\cite{Cleven:2015era, Wang:2013kra, Ma:2014zva, Hanhart:2019isz,Anwar:2021dmg,Dong:2021juy}. For simplicity, we use $D_1$ and $D_2^*$ to represent $D_1(2420)$ and $D_2^*(2460)$ in the following discussion. The effective Lagrangian is given by 
\begin{align}
{\cal L}_{\psi}=&i\frac{y_{\nr}}{2}\varepsilon^{ijk}Y^{i\dag}\left(D_{1a}^{j}\olsi D_a^{*k}-\olsi D_{1a}^j D^{*k}_a\right)+\frac{y^{\prime}_{\nr}}{\sqrt 2}\psi^{i\dag}\left(D_{2a}^{*ij}\olsi D^{*j}_a- \olsi D_{2a}^{ij} D^{*j}_a\right)+\text{h.c.},
\end{align}
where $Y^{\dag}$ ($\psi^{\dag}$) denotes the field operator that creates $\psi(4430)$ ($\psi(4415)$), $y_{\nr}$ ($y^{\prime}_{\nr}$) is the effective coupling of $\psi(4360)$ ($\psi(4415)$) to the relevant thresholds, the light flavor index $a$ runs over $u$ and $d$ quarks, and the subscript “nr” implies that the corresponding fields are normalized nonrelativistically. 
The relevant Lagrangian for the $\eta_{c1}$ and $\eta_{c1}^{\prime}$ is
\begin{align}
{\cal L}_{\eta_{c1}}=\frac{x_{\nr}^0}{\sqrt 2}\eta_{c1}^{i\dag}\left(D_{1a}^j\olsi D+\olsi D_{1a}^i \olsi D_a\right)+i\frac{x^{\prime}_{\nr}}{ 2}\varepsilon^{ijk}\eta^{\prime i\dag}_{c1}\left(D_{1a}^j\olsi D^{*k}_{a} + \olsi D^{j}_{1a} D^{*k}_{a}\right) + \text{h.c.}.
\end{align}
Here we adopt the following phase convention for the charge conjugation:
\begin{align}
{\cal C}D{\cal C}^{-1} = \olsi D, \qquad {\cal C}D^*{\cal C}^{-1} = \olsi D^*, \qquad {\cal C}D_1{\cal C}^{-1} = \olsi D_1, \qquad {\cal C}D_2^*{\cal C}^{-1} = \olsi D_2^*.
\end{align}

In accordance with the heavy quark spin symmetry (HQSS), the charmed-meson fields can be formulated in terms of the quantum numbers of the light quark degrees of freedom~\cite{Manohar:2000dt}. The two-component notation, as presented in Ref.~\cite{Hu:2005gf}, is employed under the nonrelativistic approximation for the charmed mesons. This notation defines the field of the ground state charmed mesons as 
\begin{align}
H_a=\bs P^*_a\cdot {\bs \sigma} + P_a,
\end{align}
where $\bs P^*_a$ and $P_a$ annihilates the vector and pseudoscalar charmed mesons, respectively, and $a$ represents the flavor label of the light quarks as above. 
They have quantum numbers $s_\ell^P=1/2^-$, where $s_\ell$ is the angular momentum of the light degrees of freedom and $P$ is the parity of the mesons.
For the $P$-wave charmed mesons with quantum numbers $s_\ell^P=3/2^+$, the annihilating field is 
\begin{align}
 T^i=\bs P_2^{ij}\sigma^{j}+\sqrt{\frac{2}{3}}\bs P_1^i+\sqrt{\frac{1}{6}}i\epsilon^{ijk}\bs P_1^j\sigma^k,
\end{align}
where $\bs P_2$ and $\bs P_1$ correspond to the charmed mesons $D_1$ and $D_2^*$, respectively. The fields responsible for annihilating mesons containing anticharm quarks are obtained through the transformation under charge conjugation, as detailed in Ref.~\cite{Guo:2013zbw}.

The interactions between charmed mesons and photons can be constructed based on the HQSS and the U(1) gauge invariance~\cite{Amundson:1992yp,Cheng:1992xi,Hu:2005gf,Guo:2014taa,Shi:2023mer}. The Lagrangian for the magnetic coupling of a photon to ground charmed mesons is
\begin{align}
    \mathcal{L}_{H H \gamma}
    = &\, i e \left(\beta Q_{a b}-\frac{Q^{\prime}}{m_c}\delta_{a b}\right)P_a^{* i \dagger} P_b^{* j} \left(\partial^i A^j-\partial^j A^i\right) \nonumber \\
    & +e\left(\beta Q_{a b}+\frac{Q^{\prime}}{m_c}\delta_{a b}\right) \epsilon^{i j k} \left(P_a^{* k \dagger} P_b + P_a^\dagger P_b^{*k}\right)\partial^i A^j,
\end{align}
where $Q=\text{Diag}(3/2,-1/3)$ is the light quark charge matrix, and $Q'=3/2$ is the charge of the charm quark. For a determination of the charm quark mass $m_c$ and the parameter $\beta$ using the radiative $D^*\to D \gamma$ decays, we refer to Ref.~\cite{Hu:2005gf}. Similarly, the magnetic interaction for the $P$-wave charmed mesons is
\begin{align}
    \mathcal{L}_{TT \gamma}=&\, i\left(e\beta^{\prime}Q_{ab}-\frac{eQ'\delta_{ab}}{m_c}\right)\bs P_{2a}^{im\dag}\bs P_{2b}^{ni}(\partial^mA^n-\partial^nA^m\nonumber)\nonumber\\
    &+\left(\frac{e\beta^{\prime}Q_{ab}}{\sqrt 6}+\sqrt{\frac{3}{2}}\frac{eQ^{\prime}\delta_{ab}}{m_c}\right)\varepsilon^{lmn}\left(\bs P_{2a}^{il\dag}\bs P_{1b}^i+\bs P_{1a}^{i\dag}\bs P_{2a}^{il}\right)\partial^mA^n\nonumber\\
    &+i\left(\frac{5e\beta^{\prime}Q_{ab}}{6}+\frac{eQ^{\prime}\delta_{ab}}{2m_c}\right)\bs P_{1a}^{m\dag}\bs P_{1b}^n(\partial^mA^n-\partial^nA^m).
    \label{Eq:Lag_TTgamma}
\end{align}
with a parameter $\beta^{\prime}$. In our calculation, the parameter $\beta$ is determined from the radiative decay of $D^*\to D \gamma$~\cite{Hu:2005gf} from experimental data~\cite{ParticleDataGroup:2024cfk}, while $\beta^{\prime}$ can be determined from the decay width for $D_2^*\to D_1\gamma$ using quark-model predictions in Ref.~\cite{Goity:2000dk}.

\subsection{Amplitudes for the radiative decays of $\psi(4360)$ and $\psi(4415)$}

The $\eta_{c1}$ and $\eta_{c1}^{\prime}$ can be produced through the radiative decays of $\psi(4360)$ and $\psi(4415)$, as depicted in Fig.~\ref{Fig:decay_psi}. The decay amplitudes for $\psi^i(4360)(p)\to \gamma^j(q) \eta_{c1}^k(p-q)$ and $\psi^i(4415)(p)\to \gamma^j(q) \eta_{c1}^{\prime k}(p-q)$ are given by
\begin{align}
i{\cal M}_{Y}(p)&=2N_1\frac{i g_{\nr} \chi_{\nr}}{2\sqrt{2}} \left(\frac{e\beta}{3} +\frac{4e}{3m_c}\right)\left[\boldsymbol q\cdot \varepsilon(\eta_{c1})\varepsilon(\gamma)\cdot \varepsilon^*(Y)-\boldsymbol q\cdot \epsilon^*(Y)\varepsilon(\gamma)\cdot \varepsilon(\eta_{c1})\right] I(\bs q),\label{Eq:decay_amp_Y}\\
i{\cal M}_{\psi}(p)&=2N_2\frac{5}{6}\frac{ig^{\prime}_{\nr}\chi^{\prime}_{\nr}}{2\sqrt 2}\left(\frac{e\beta'}{3\sqrt 6}+\sqrt{\frac{3}{2}}\frac{4e}{3m_c^{\prime}}\right)\left[\boldsymbol q\cdot\varepsilon^{*}(\psi)\varepsilon(\gamma)\cdot\varepsilon(\eta^{\prime}_{c1})-\varepsilon^{*}(\psi)\cdot\varepsilon(\gamma)\boldsymbol q\cdot\varepsilon(\eta^{\prime}_{c1})\right]I(\bs q),
\label{Eq:decay_amp_psi}
\end{align}
where $N_1=8m_{D^*}m_{D_1}m_{D}(m_Ym_{\eta_{c1}})^{1/2}$ and $N_2=8m_{D_2}m_{D_1}m_{D^*}(m_{\psi}m_{\eta^{\prime}_{c1}})^{1/2}$ account for the nonrelativistic normalization, $\bs q$ is the three-momentum of the photon, and the factor of $2$ arises from the contribution of charge conjugation parts. Here, to ease the notation, we have used $Y$ and $\psi$ to represent $\psi(4360)$ and $\psi(4415)$, respectively. $I(\bs q)$ represents the scalar three-point loop function 
\begin{align}
I(\bs q)=i\int\frac{d^4l}{(2\pi)^4}\frac{1}{l^2-m_1^2+im_1\Gamma_1}\frac{1}{(p-l)^2-m_2^2+im_2\Gamma_2}\frac{1}{(l-q)^2-m_3^2+im_3\Gamma_3},
\label{Eq:triangle_loop}
\end{align}
where the subscripts on the meson masses $m_i$ and their decay widths $\Gamma_i$, $i=1,2,3$, refer to the mesons $D^*$, $\olsi D_1$, and $D$ for the amplitude ${\cal M}_Y(p)$, and $D_2^*$, $\olsi D^*$, and $D_1$ for the amplitude ${\cal M}_\psi(p)$. The decay widths of $D_1$ and $D_2^*$ are considered using complex masses, and the loop function is numerically computed using the \texttt{LoopTools} package~\cite{Hahn:1998yk}.

\begin{figure}[tb]
\centering	
\includegraphics[width = 0.34\textwidth
]{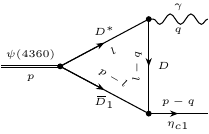} \hspace{1cm}
\includegraphics[width = 0.34\textwidth]{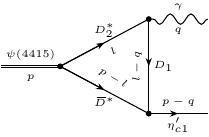}\hfill
\caption{Feynman diagrams for the decays $\psi(4360)\to\gamma \eta_{c1}$ and $\psi(4415)\to\gamma \eta_{c1}^{\prime}$. The charge-conjugated diagrams are not shown. }
\label{Fig:decay_psi}
\end{figure}
Since the initial and final states of the two decays have the same quantum numbers, the decay amplitudes for $\psi(4360)\to\gamma\eta_{c1}$ and $\psi(4415)\to\gamma\eta^{\prime}_{c1}$ exhibit the same tensor structure, specifically $\boldsymbol q\cdot\varepsilon^{*}(\psi)\varepsilon(\gamma)\cdot\varepsilon(\eta^{\prime}_{c1})-\varepsilon^{*}(\psi)\cdot\varepsilon(\gamma)\boldsymbol q\cdot\varepsilon(\eta^{\prime}_{c1})$. This structure ensures the gauge invariance of the amplitudes. 
As detailed in Eq.~\eqref{Eq:polarization}, the angular distributions are derived from this tensor structure as
\begin{align}
\frac{d\Gamma}{d~\text{cos}~\theta}\propto \pi|\bs q|^2\left(4+2\,\text{sin}^2~\theta\right),
\label{Eq:Lorentz_str}
\end{align}
where $\theta$ is the angle between the $z$-axis (beam axis) and $\bs q$.

\subsection{Input parameters}

For an $S$-wave shallow bound state, the coupling of this state to its constituent hadrons can be related to its binding energy. The effective coupling with nonrelativistic normalization is~\cite{Weinberg:1965zz,Baru:2003qq,Cleven:2011gp}
\begin{align}
g_{\nr}^2=\lambda^2\frac{2\pi}{\mu^2}\gamma\left[1+{\cal O}\left(\frac{\gamma}{\beta}\right)\right],
\label{Eq:coupling}
\end{align}
where the binding momentum is $\gamma=\sqrt{2\mu E_B}$ with the reduced mass $\mu=m_1m_2/(m_1+m_2)$ and the binding energy $E_B$. 
The hard scale $\beta$ is a hadronic scale associated with the inverse range of the forces. For the systems under study, the binding momenta are smaller than $0.4$~GeV and thus much smaller than the involved charmed meson masses; $\beta$ can be estimated by the masses of the light vector mesons, such as $m_\rho$, that are assumed to play a dominant role in providing the attraction forming the bound states~\cite{Dong:2021juy}. 
The parameter $\lambda^2$ represents the compositeness of the state, with $\lambda^2=1$ for a pure hadronic molecule. 
In this work, we employ the isospin-averaged masses for the $D$, $D^*$, $D_1$, and $D_2^*$ mesons. 
We consider $\psi(4360)$ and $\psi(4415)$ as pure isoscalar vector molecules of $D^*\olsi D_1$ and $D^*\olsi D_2^*$, respectively, with masses $M_{Y}=(4374\pm 7)$~MeV and $M_{\psi}=(4415\pm 5)$~MeV~\cite{ParticleDataGroup:2024cfk}, respectively. Then we estimate the effective couplings $y_{\nr}$ and $y^{\prime}_{\nr}$ in Eqs.~\eqref{Eq:decay_amp_Y} and~\eqref{Eq:decay_amp_psi} to be
\begin{align}
\left|y_{\nr}\right|=(1.36\pm 0.04)~\text{GeV}^{-1/2}, \qquad \left|y^{\prime}_{\nr}\right|=(1.34\pm 0.03)~ \text{GeV}^{-1/2},
\label{Eq:couple_psi}
\end{align}
which exhibit a good HQSS behavior.
Since the values of the effective couplings $y_{\nr}$ and $y^{\prime}_{\nr}$ are almost identical in the pure hadronic molecular picture,
we also assume $x_{\nr}\simeq x_{\nr}^{\prime}$. Taking the unquenched lattice QCD result for the lowest $1^{-+}$ charmonium-like state mass as input~\cite{HadronSpectrum:2012gic}, the binding energy for $\eta_{c1}$ is $(72\pm 16)$~MeV, and we assume that the binding energy for $\eta_{c1}^{\prime}$ is the same. The relevant effective couplings are then estimated to be
\begin{align}
\left|x_{\nr}\right|=\left|x^{\prime}_{\nr}\right|=(1.96\pm0.03)~ \text{GeV}^{-1/2}.
\label{Eq:couple_eta}
\end{align}
Note that the above effective couplings bear another relative uncertainty of about $\gamma/\beta\sim 50\%$ due to the sizable binding momenta.

The charm quark mass $m_c$ and the parameter $\beta$ in the magnetic coupling of the $S$-wave charmed meson are estimated using the radiative decay of $D^*\to D\gamma$. Specifically, we adopt the values $m_c=1.5$~GeV and $\beta^{-1}=276$~MeV, as reported in Ref.~\cite{Hu:2005gf}. For the radiative decay of $P$-wave charmed mesons, we rely on the quark model prediction for the width of $D_2^*\to D_1\gamma$ to determine the value of $\beta^{\prime}$, given the absence of experimental measurements.
The radiative decay width for $D_2^*(p)\to \gamma(q) D_1(k)$ is given by
\begin{align}
\Gamma_{D_2^*}=\left(\frac{e\beta'Q_{ab}}{\sqrt 6}+\sqrt{\frac{3}{2}}\frac{2e}{3m_c}\right)^2\frac{m_{D_1}\left|\bs q\right|^3}{12\pi m_{D^*_2}}.
\label{Eq:Decay_D2}
\end{align}
With $\left|\bs q\right|=32$~MeV and $m_c=1.5$~GeV, the predicted widths for $D_2^{*-}\to \gamma D_1^-$ and $\olsi D_2^{*0}\to \gamma \olsi D_1^0$ are $0.1_{-0.1}^{+0.4}$ and $180_{-171}^{+576}$ eV, respectively~\cite{Goity:2000dk}.
The huge difference is due to the difference in the interference between the two terms in the bracket.
By fitting these decay widths, we determine $\beta^{\prime-1}=235^{+8}_{-6}$ MeV.

\section{Radiative decays of $\psi(4360)$ and $\psi(4415)$}\label{Sec:decay_width}

The radiative decay widths for $\psi(4360)\to\gamma\eta_{c1}$ and $\psi(4415)\to\gamma\eta_{c1}^{\prime}$ are estimated using the triangle loops, as depicted in Fig.~\ref{Fig:decay_psi}. 
Due to the absence of experimental measurements for the $\eta_{c1}$ and $\eta_{c1}^{\prime}$, we fix the masses of $\psi(4360)$ and $\psi(4415)$~\cite{ParticleDataGroup:2024cfk}, while varying the binding energy of $\eta_{c1}$ and $\eta_{c1}^{\prime}$ within the range of [0, 100]~MeV (so that the effective couplings change according to Eq.~\eqref{Eq:coupling}). The resulting partial decay widths are illustrated in Fig.~\ref{Fig:decay_Eb}. 
Notably, the radiative decay width of $\psi(4415)$ is more than one order of magnitude smaller than that of $\psi(4360)$ when the binding energies of $\eta_{c1}$ and $\eta_{c1}^{\prime}$ are the same. This difference arises because the decay widths are proportional to $|\bs q|^3$, and the small relative momentum $\bs q$ in the decay of $\psi(4415)$ suppresses its width. 
A similar suppression leads to a small decay width for $\psi(4360)\to \gamma\eta_{c1}$ when compared to the radiative decay $Y(4260)\to \gamma X(3872)$ calculated in Ref.~\cite{Guo:2013zbw}. Furthermore, due to the quick change of the phase space, for $\eta_{c1}^{\prime}$ with a binding energy $E_B=54$~MeV, the partial width is $\Gamma[\psi(4415)\to \gamma\eta^{\prime}_{c1}]=0.21$~keV, which is only $10\%$ of the width for $E_B=100$~MeV, and the decay width becomes just a few eV when $E_B<30$~MeV. This result indicates that, once $\eta_{c1}^{\prime}$ is a shallow $D^*\olsi D_2^*$ bound state, it will be challenging to observe $\eta_{c1}^{\prime}$ in the radiative decay of $\psi(4415)$, at least at BESIII.

\begin{figure}[tb]
\centering	
\includegraphics[scale=0.8
]{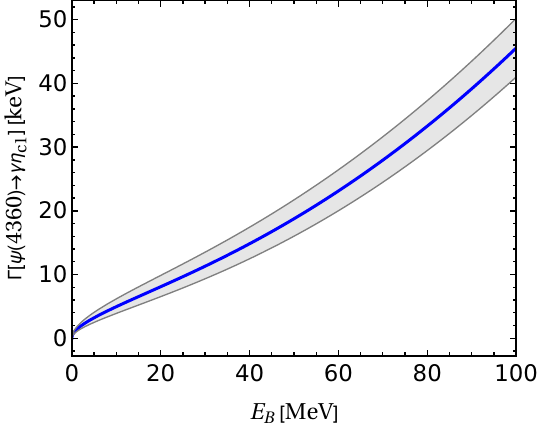}
\includegraphics[scale=0.84]{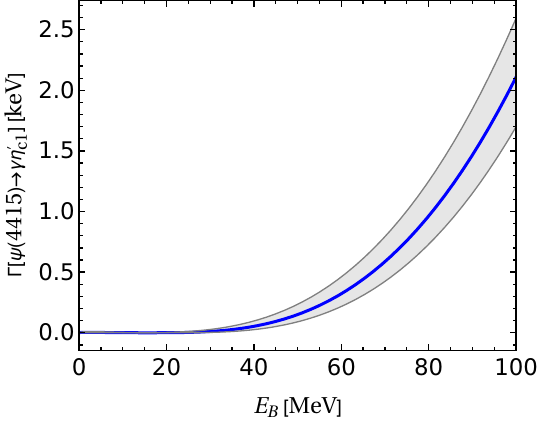}\hfill
\caption{Decay widths for $\psi(4360)\to \gamma \eta_{c1}$ and $\psi(4415)\to \gamma \eta_{c1}^{\prime}$. The variable $E_B$ represents the binding energy of $\eta_{c1}$ or $\eta_{c1}^{\prime}$. The gray bands reflect the experimental uncertainties associated with the masses of $\psi(4360)$ and $\psi(4415)$~\cite{ParticleDataGroup:2024cfk}.
\label{Fig:decay_Eb}}
\end{figure}

\begin{figure}[tb]
\centering	
\includegraphics[scale=0.8
]{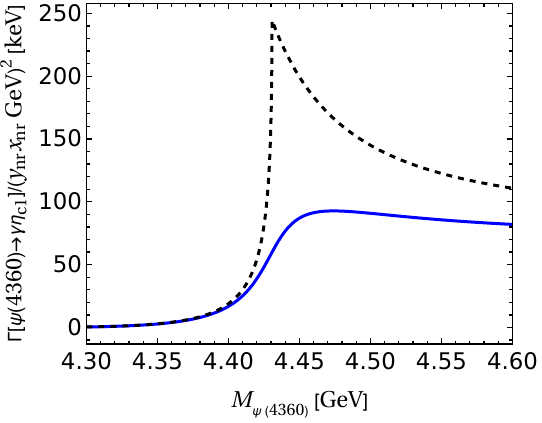}
\includegraphics[scale=0.8]{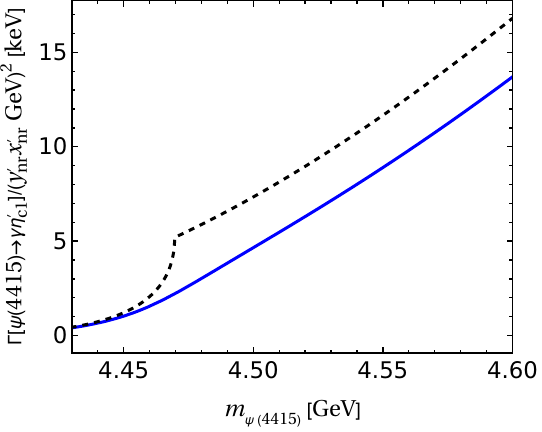}\hfill
\caption{Radiative decay widths of $\psi(4360)$ and $\psi(4415)$ as functions of their masses. The blue and black lines denote that the decay widths of $D_1$ and $D_2^*$ are included and excluded, respectively, in our calculation.
\label{Fig:decay_Mpsi}}
\end{figure}

\begin{figure}[tb]
    \centering	
    \includegraphics[scale=0.8
    ]{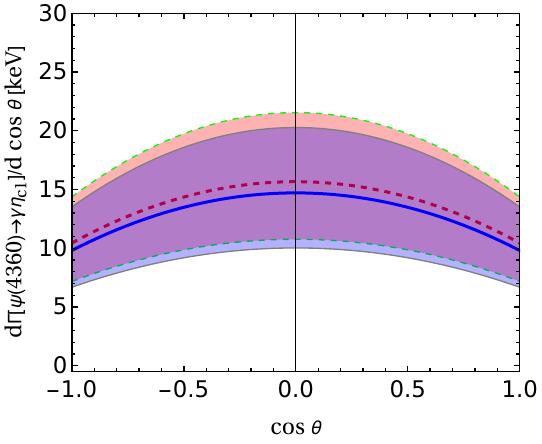}
    \includegraphics[scale=0.8]{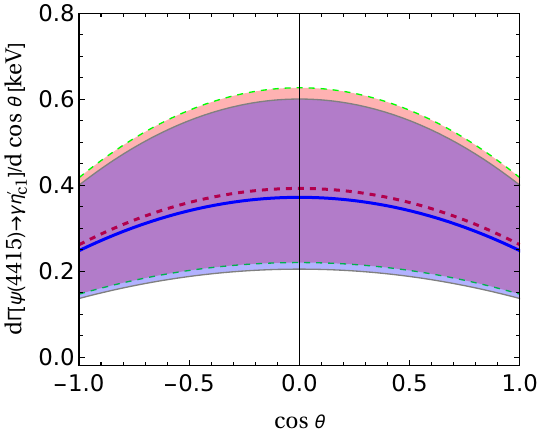}\hfill
    \caption{Angular distributions for the $\psi(4360)\to\gamma \eta_{c1}$ and $\psi(4415)\to\gamma \eta_{c1}'$. Here, $\theta$ refers to the angle between the beam axis and the momentum of the emitted photon in the rest frame of the initial state. The blue solid and red dashed lines denote the differential widths with and without including the decay widths of $D_1$ and $D_2^*$, respectively. The errors in the distribution account for the uncertainties in the masses of $\psi(4360)$, $\psi(4415)$, $\eta_{c1}$, and $\eta_{c1}^{\prime}$. The binding energy of $\eta_{c1}$ and $\eta_{c1}^{\prime}$ is taken to be $72(16)$~MeV from the lattice QCD calculations in Ref.~\cite{HadronSpectrum:2012gic}. 
    \label{Fig:angle_dist}}
\end{figure}

The $1^{--}$ charmonium-like state $\psi$ can be produced in $e^+e^-$ annihilations through the process $e^+e^-\to \gamma^{*}\to \psi$. To investigate the variation of the radiative decay width as a function of the initial energy, we modify the c.m. energy $\sqrt s$.
As depicted in Fig.~\ref{Fig:decay_Mpsi}, we vary the masses of $\psi(4360)$ and $\psi(4415)$, corresponding to $\sqrt s$ in $e^+e^-$ annihilations, from the thresholds of $D^*\olsi D_1$ and $D^*\olsi D_2^*$ up to $4.6$~GeV, and evaluate the relative decay widths. 
To show more clearly the origin of nontrivial energy dependence of the radiative decay widths, we adopt two approaches: one including and the other excluding the decay widths of $D_1$ and $D_2^*$ in the triangle loops illustrated in Fig.~\ref{Fig:decay_psi}. 
It is worth noting that the effective couplings have been normalized in the plots, ensuring that the curves solely reflect the behavior of the triangle loops and kinematics and are not affected by the large uncertainty of the effective couplings.
When the decay widths of the intermediate $D_1$ and $D_2$ mesons are neglected, cusps, as shown in the black dashed lines in Fig.~\ref{Fig:decay_Mpsi}, appear at the $D^*\olsi D_1$ and $D^*\olsi D_2^*$ thresholds.  
The left panel in Fig.~\ref{Fig:decay_Mpsi} exhibits a more pronounced cusp structure compared to the right panel, which is attributed to the difference in the distances between the triangle singularities (TSs)~\cite{Guo:2019twa} of the triangle diagrams in Fig.~\ref{Fig:decay_psi} and their respective thresholds ($D_1\olsi D^*$ and $D_2\olsi D^*$ for the left and right panels, respectively). 
With the binding energies of $\eta_{c1}$ and $\eta_{c1}'$ set to 72~MeV, the TS for the left plot in Fig.~\ref{Fig:decay_psi} is located at $4.36-i0.03$~GeV, significantly closer to the $D_1\olsi D^*$ threshold than the TS for the right plot, located at $4.40-i0.01$~GeV, is to the $D_2\olsi D^*$ threshold.
Inclusion of the $D_1$ and $D_2$ widths in the triangle loops smears the cusp effects and substantially suppresses the relevant partial widths. For the decay $\psi(4360)\to\gamma\eta_{c1}$, the peak of the decay width shifts from the $D^*\olsi D_1$ threshold at 4.43~GeV to $4.47$~GeV.

Furthermore, Fig.~\ref{Fig:angle_dist} presents the predicted angular distributions in Eq.~\eqref{Eq:Lorentz_str} for the radiative decays of $\psi(4360)$ and $\psi(4415)$, where $72(16)$~MeV was used as input for the binding energies of both $\eta_{c1}$ and $\eta_{c1}'$. The minimum of the differential width differs from the maximum by approximately $27\%$ for both the $\psi(4360)\to \gamma \eta_{c1}$ and $\psi(4415)\to \gamma\eta_{c1}^{\prime}$ processes. 

\section{Production of ${\eta_{c1}}$ and ${\eta_{c1}^{\prime}}$ in ${e^+e^-}$ collisions}\label{Sec:production}

The production of $1^{-+}$ exotic states in $e^+e^-$ annihilations proceeds through the processes $e^+e^-\to\gamma^{*}\to \psi(4360)\to\gamma\eta_{c1}$ and $e^+e^-\to\gamma^{*}\to \psi(4415)\to\gamma\eta_{c1}^{\prime}$. The production mechanism for $\eta_{c1}$ and $\eta_{c1}^{\prime}$ is similar to that of $X(3872)$, suggested in Ref.~\cite{Guo:2013zbw} and later confirmed by the BESIII Collaboration~\cite{BESIII:2013fnz}.
By utilizing the cross section for $e^+e^-\to \gamma X(3872)$ measured by the BESIII Collaboration and the decay width for $\psi(4415)\to e^+e^-$, we estimate the cross sections for $e^+e^-\to \psi(4360)\to \gamma \eta_{c1}$ and $e^+e^-\to \psi(4415)\to\gamma \eta_{c1}^{\prime}$.

In the vicinity of the mass regions of $\psi(4360)$ and $\psi(4415)$, the coupling between the photon $\gamma^{*}$ and $\psi$ (where $\psi$ represents either $\psi(4360)$ or $\psi(4415)$) can be estimated using the VMD model assuming the production proceeds mainly through the intermediate $\psi$ states.
The amplitude for the production of the molecule $X$ is then given by
\begin{align}
M_{\text{prod}}=-ie^2\olsi u(p_1)\gamma_{\mu}v(p_2) \frac{Q_cf_{\psi}}{m_{\psi}}\left(g_{\mu\nu}-\frac{p_{\mu} p_{\nu}}{m_{\psi}^2}\right)\frac{-i\varepsilon_\nu(\psi)}{(p^2-m_{\psi}^2+im_{\psi}\Gamma^{\prime}_{\psi})}{\cal M}_{\psi}(p),
\label{Eq:amp_production}
\end{align}
where $Q_c=2/3$ is the charge of the charm quark, $p=p_1+p_2$ is the total momentum of $e^+e^-$, $s=p^2$, $\Gamma^{\prime}_{\psi}$ is the experimental decay width of $\psi$, and ${\cal M}_{\psi}(p)$ is the amplitude for the radiative decay $\psi\to\gamma X$ with $X$ representing either $\eta_{c1}$ or $\eta_{c1}^{\prime}$.  
We use constant decay widths for the $\psi$ states, which is sufficient for providing order-of-magnitude estimates. 
By neglecting the masses of the electron and positron, the cross section can be simplified to
\begin{align}
\sigma_{\psi}
=\frac{e^4f_{\psi}^2}{9m_{\psi}^2\sqrt s}\frac{\Gamma_{\psi\to\gamma X}(\sqrt s)}{\left|\sqrt s - m_{\psi} +i\Gamma_{\psi}'/2\right|^2},
\label{Eq:cross_section}
\end{align}
where $\Gamma_{\psi\to \gamma X}(\sqrt s)$ represents the partial width for the production of the molecule in the radiative decay of $\psi\to \gamma X$.

The parameter $f_{\psi}$ for $\psi(4360)$ can be determined from experimental measurements. The BESIII Collaboration has measured an upper limit of the Born cross section at $90\%$ confidence level, specifically $\sigma^B[e^+e^-\to\gamma X(3872)]\cdot {\cal B}[X(3872)\to\pi^+\pi^-J/\psi]=(0.11\pm0.09\pm0.01)$~pb at $\sqrt s=4.360$~GeV~\cite{BESIII:2013fnz}. 
The branching ratio ${\cal B}[X(3872)\to\pi^+\pi^-J/\psi]$ is estimated to be $\sim 6\%$ in Ref.~\cite{Ji:2025hjw}.
With this, we can determine  $f_{\psi(4360)}$ using the cross section in Eq.~\eqref{Eq:cross_section} and the radiative decay width of $\psi(4360)\to\gamma X(3872)$ provided in Appendix~\ref{Sec:decay_X3872}. 
By accounting for the uncertainties in the Born cross section, we estimate $f_{\psi(4360)}$ to be $0.74_{-0.52}^{+0.28}$~GeV. Furthermore, the decay constant $f_{\psi(4415)}$ is derived from its partial width for the $\psi(4415)\to e^+e^-$ process~\cite{ParticleDataGroup:2024cfk}. 
Employing the formula from our previous work~\cite{Shi:2023ntq}, we obtain $f_{\psi(4415)}=(0.16\pm0.01)$~GeV. 
Notably, despite the significant difference between the central values of the two decay constants, the value of $f_{\psi(4360)}$ agrees with that of $f_{\psi(4415)}$ within $1.1\sigma$. 

\begin{figure}[tb]
    \centering	
    \includegraphics[scale=0.8
    ]{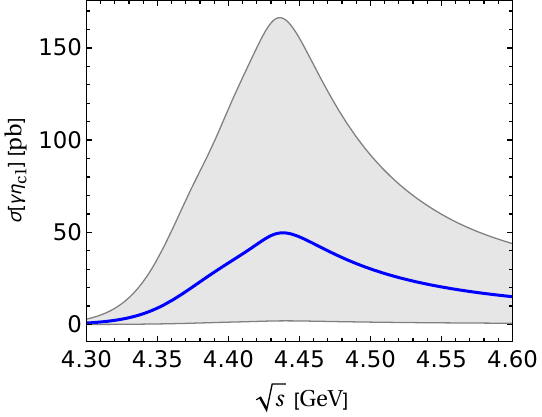}
    \includegraphics[scale=0.8]{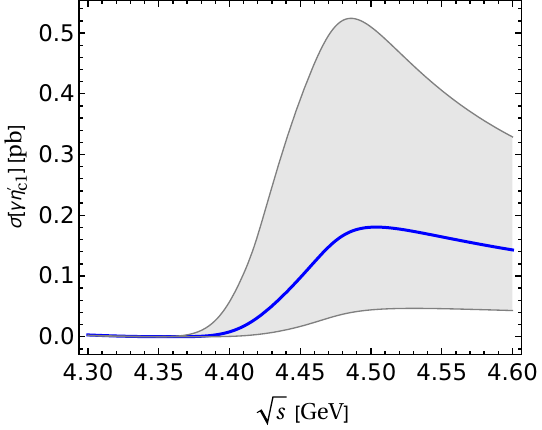}\hfill
    \caption{Cross sections for the production of $\eta_{c1}$ and $\eta_{c1}^{\prime}$ in $e^+e^-$ annihilations. The left and right panels show the cross sections for $e^+e^-\to \psi(4360)\to\gamma\eta_{c1}$ and $e^+e^-\to \psi(4415)\to\gamma\eta_{c1}^{\prime}$, respectively, as the c.m. energy $\sqrt s$ increases.
    The gray bands represent the total uncertainties in the cross sections, obtained by adding in quadrature the contributions from various parameter uncertainties, including those of the masses and decay widths of $\psi(4360)$ and $\psi(4415)$, the decay constant $f_{\psi}$ in Eq.~\eqref{Eq:cross_section}, the parameter $\beta^{\prime}$ in Eq.~\eqref{Eq:Decay_D2}, the binding energies of $\eta_{c1}$ and $\eta^{\prime}_{c1}$, and the higher-order $\mathcal{O}(\gamma/\beta)$ term in Eq.~\eqref{Eq:coupling}.
    \label{Fig:production_Mpsi}}
\end{figure}

The cross sections for the production of $\eta_{c1}$ and $\eta_{c1}^{\prime}$ in $e^+e^-$ annihilations are then estimated, as shown in Fig.~\ref{Fig:production_Mpsi}. At $\sqrt s =4.44$~GeV, the cross section for $e^+e^-\to\psi(4360)\to \gamma\eta_{c1}$ reaches its maximum value, which is $\sigma^{\text{max}}[\gamma\eta_{c1}]=(50_{-48}^{+116})$~pb, while the cross section is $\sigma[\gamma\eta_{c1}]=(13_{-13}^{+38})$~pb at $\sqrt s=4.36$~GeV.
Here the uncertainties result from the combined effect of various parameter uncertainties added in quadrature, including those from the masses and decay widths of $\psi(4360)$ and $\psi(4415)$, the decay constant $f_{\psi}$ in Eq.~\eqref{Eq:cross_section}, the parameter $\beta^{\prime}$ in Eq.~\eqref{Eq:Decay_D2}, the binding energies of $\eta_{c1}$ and $\eta^{\prime}_{c1}$, and the higher-order $\mathcal{O}(\gamma/\beta)$ term in Eq.~\eqref{Eq:coupling}.
Compared to the maximum decay width of $\psi(4360)$ shown in Fig.~\ref{Fig:decay_Mpsi}, the energy $\sqrt s$ corresponding to the maximum cross section decreases by tens of MeV. 
For the production of $\eta_{c1}^{\prime}$ in the process $e^+e^-\to\psi(4415)\to\gamma\eta_{c1}^{\prime}$, the maximum cross section is $\sigma^{\text{max}}[\gamma\eta_{c1}^{\prime}]=(0.18_{-0.14}^{+0.32})$~pb at $\sqrt s=4.50$~GeV, while $\sigma[\gamma\eta_{c1}^{\prime}]=(24_{-~21}^{+109})$~fb at $\sqrt s =m_{\psi(4415)}$. This indicates that the optimal energy region for the observation of $\eta_{c1}$ and $\eta_{c1}^{\prime}$ is located above the masses of $\psi(4360)$ and $\psi(4415)$, respectively.
In comparison to the previous calculation in Ref.~\cite{Wang:2014wga}, our prediction for the cross section of $e^+e^-\to\psi(4360)\to \gamma\eta_{c1}$ is significantly larger. Notably, at $\sqrt s=4.36$~GeV, our result exceeds the one presented in Ref.~\cite{Wang:2014wga} by approximately one order of magnitude. Moreover, the energy $\sqrt s$ corresponding to the maximum cross section is also higher than that predicted in the previous calculation.

To provide theoretical guidance for the search of $\eta_{c1}$ and $\eta_{c1}^{\prime}$, we estimate the number of events that can be generated in $e^+e^-$ annihilations. 
From 2011 to 2014, the BESIII experiment accumulated an integrated luminosity of 47 and 112~pb$^{-1}$ at $\sqrt s=4.42$~GeV and $4.53$~GeV, respectively~\cite{BESIII:2015qfd,BESIII:2022dxl}. 
Based on our predictions, we conservatively estimate that at least about 79~events of $\eta_{c1}$, using the lower limit of the cross section in the left panel of Fig.~\ref{Fig:production_Mpsi}, and about 19~events of $\eta_{c1}^{\prime}$ could have been produced at BESIII during this period. 
However, since the $\eta_{c1}$ and $\eta_{c1}^{\prime}$ need to be reconstructed in final states such as $J/\psi\gamma$, $\eta_c\eta$ and $D\bar D^{(*)}\pi$, it could be difficult to observe the $\eta_{c1}'$ at BESIII.
With the high luminosity of SCTF (approximately 1~ab$^{-1}$/year), we predict that at least ${\cal O}(10^6)$ events of $\eta_{c1}$ can be generated annually at $\sqrt s\approx4.44$~GeV, and roughly ${\cal O}(10^5)$ events of $\eta_{c1}^{\prime}$ can be produced at $\sqrt s\approx4.50$~GeV. Consequently, we anticipate that both the $\eta_{c1}$ and $\eta_{c1}^{\prime}$ can be observed at the SCTF. 

\section{Conclusion}\label{Sec:conlusion}

The $e^+e^-$ annihilation process plays a crucial role in the search for exotic states. Specifically, charmonium-like states with exotic $J^{PC}=\text{odd}^{-+}$ quantum numbers can be searched for through the reaction $e^+e^-\to\psi\to\gamma X$ involving the $1^{--}$ charmonium-like $\psi$ states. 
In this study, we calculate the decay widths for the processes $\psi(4360)\to\gamma\eta_{c1}$ and $\psi(4415)\to\gamma\eta_{c1}^{\prime}$ under the assumption that $\psi(4360)$ and $\psi(4415)$ are $D^*\olsi D_1$ and $D^*\olsi D_2^*$ molecules, respectively. 
Utilizing the measurement of $e^+e^-\to\gamma X(3872)$ at $\sqrt s=4.36$~GeV~\cite{BESIII:2013fnz} and the partial width for $\psi(4415)\to e^+e^-$~\cite{ParticleDataGroup:2024cfk}, we estimate the cross sections for the processes $e^+e^-\to\psi(4360)\to\gamma\eta_{c1}$ and $e^+e^-\to\psi(4415)\to\gamma\eta_{c1}^{\prime}$. 
Our results indicate that the cross sections for the production of $\eta_{c1}$ and $\eta_{c1}^{\prime}$ are of the order of ${\cal O}(10)$~pb and ${\cal O}(0.01)$~pb, respectively. 
Our calculation of the cross sections at different c.m. energies $\sqrt s$ has revealed that the value of $\sqrt s$ corresponding to the maximum cross section is affected by the singularities present in the triangle loops, specifically the triangle singularity and threshold cusp. 
Based on these findings, we suggest searching for $\eta_{c1}$ and $\eta_{c1}^{\prime}$ at high-luminosity $e^+e^-$ annihilation experiments, such as the future STCF, at $\sqrt s\approx4.44$~GeV and $4.50$~GeV, respectively.

\vspace{-10pt}

\begin{acknowledgments}
We are grateful to Dian-Yong Chen for useful comments.
This work is supported in part by the National Natural Science Foundation of China (NSFC) under Grants No. 12361141819, No. 12125507,  and No. 12047503; by the Chinese Academy of Sciences (CAS) under Grant No. YSBR-101; by the National Key R\&D Program of China under Grant No. 2023YFA1606703; and by MICIU/AEI/10.13039/501100011033 under contracts No. PID2020-112777GB-I00. P.-P. Shi thanks the Institute of Theoretical Physics, CAS, for the hospitality during his visit. 
\end{acknowledgments}

\appendix

\section{Polarization vector}\label{Sec:polarization}

In the nonrelativistic limit, the polarization vectors along the $\bs q$ direction are
\begin{align}
 \varepsilon(0)=\begin{pmatrix}
\sin \theta \cos \phi \\
\sin \theta \sin \phi \\
\cos \theta
\end{pmatrix},
\qquad \varepsilon(\pm 1)=\frac{1}{\sqrt{2}}\begin{pmatrix}
\cos \theta \cos \phi\mp i \sin \phi \\
\cos \theta \sin \phi\pm i\cos \phi  \\
- \sin \theta
\end{pmatrix},\label{Eq:polarization}
\end{align}
where $\theta$ and $\phi$ are the polar and azimuthal angles of $\bs q$ with respect to the $z$-axis, respectively.
With these polarization vectors, the modulus squared of the tensor structure in Eqs.~\eqref{Eq:decay_amp_Y} and \eqref{Eq:decay_amp_psi} is reduced to
\begin{align}
&\int_{0}^{2\pi}d\phi \sum_{\text{polarizations}}\left|\bs q\cdot\varepsilon^{*}(\psi)\varepsilon(\gamma)\cdot\varepsilon(\eta^{\prime}_{c1})-\varepsilon^{*}(\psi)\cdot\varepsilon(\gamma)\bs q\cdot\varepsilon(\eta^{\prime}_{c1})\right|^2 \nonumber\\
=&\int_0^{2\pi}d\phi\sum_{\text{polarizations}}\Big[\bs q_i\varepsilon_i^*(\psi)\varepsilon_j(\gamma)\varepsilon_j(\eta_{c1}^{\prime})\bs q_{i^{\prime}}\varepsilon_{i^{\prime}}(\psi)\varepsilon^*_{j^{\prime}}(\gamma)\varepsilon^*_{j^\prime}(\eta_{c1}^{\prime})\nonumber\\
&-\bs q_i\varepsilon_i^*(\psi)\varepsilon_j(\gamma)\varepsilon_j(\eta_{c1}^{\prime})\bs q_{i^{\prime}}\varepsilon_{j^{\prime}}(\psi)\varepsilon^*_{j^{\prime}}(\gamma)\varepsilon^*_{i^\prime}(\eta_{c1}^{\prime})
-\bs q_i\varepsilon_j^*(\psi)\varepsilon_j(\gamma)\varepsilon_i(\eta_{c1}^{\prime})\bs q_{i^{\prime}}\varepsilon_{i^{\prime}}(\psi)\varepsilon^*_{j^{\prime}}(\gamma)\varepsilon^*_{j^\prime}(\eta_{c1}^{\prime})\nonumber\\
&+\bs q_i\varepsilon_j^*(\psi)\varepsilon_j(\gamma)\varepsilon_i(\eta_{c1}^{\prime})\bs q_{i^{\prime}}\varepsilon_{j^{\prime}}(\psi)\varepsilon^*_{j^{\prime}}(\gamma)\varepsilon^*_{i^\prime}(\eta_{c1}^{\prime})
\Big]\nonumber\\
=&\int_0^{2\pi}d\phi\sum_{\lambda=\pm 1}\Big[
\bs q_i \bs q_{i^{\prime}}\delta_{jj^{\prime}}\left(\delta_{ii^{\prime}}-\delta_{i3}\delta_{i^{\prime}3}\right)\varepsilon_{j}(\gamma,\lambda)\varepsilon_{j^\prime}^*(\gamma,\lambda)
-\bs q_i \bs q_{i^{\prime}}\delta_{ji^{\prime}}\left(\delta_{ij^{\prime}}-\delta_{i3}\delta_{j^{\prime}3}\right)\varepsilon_{j}(\gamma,\lambda)\varepsilon_{j^\prime}^*(\gamma,\lambda)\nonumber\\
&-\bs q_i \bs q_{i^{\prime}}\delta_{ij^{\prime}}\left(\delta_{ji^{\prime}}-\delta_{j3}\delta_{i^{\prime}3}\right)\varepsilon_{j}(\gamma,\lambda)\varepsilon_{j^\prime}^*(\gamma,\lambda)
+\bs q_i \bs q_{i^{\prime}}\delta_{ii^{\prime}}\left(\delta_{jj^{\prime}}-\delta_{j3}\delta_{j^{\prime}3}\right)\varepsilon_{j}(\gamma,\lambda)\varepsilon_{j^\prime}^*(\gamma,\lambda)
\Big]\nonumber\\
=& \pi|\bs q|^2\left(4+2\,\text{sin}^2\theta\right),
\end{align}
where the subscripts $i^{(\prime)}$ and $j^{(\prime)}=1,2,3$ denote the spatial components, $\sum\epsilon_i(\psi)\epsilon^*_{i^\prime}(\psi)=\delta_{ii^\prime}-\delta_{i3}\delta_{i^{\prime}3}$, and $\sum\epsilon_j(\eta_{c1}^{\prime})\epsilon^*_{j^\prime}(\eta_{c1}^{\prime})=\delta_{jj^\prime}$.
Here, the photon polarizations are only explicitly specified after the second equality sign, and we use the polarization vectors of the photon as defined in Eq.~\eqref{Eq:polarization} to derive the final expression.

\section{Decay width for $\psi(4360)\to\gamma X(3872)$}\label{Sec:decay_X3872}

The E1 transition between the $P$- and $S$-wave charmed mesons is parameterized as~\cite{Guo:2013zbw}
\begin{align}
{\cal L}_{TH\gamma}=\sum_a\frac{c_a}{2}\Tr[T^i_aH^{\dag}_a]E^i+\text{h.c.},
\end{align}
where the subscript $a=(u,d)$ is the light-flavor index. Due to the lack of experimental measurements, the values of the effective coupling $c_a$ are extracted from the radiative decay widths for $D_1^{0(+)}\to D^{0(+)}\gamma$ estimated in quark models~\cite{Fayyazuddin:1994qu,Godfrey:2005ww}. The values of $c_u$ and $c_{d}$ are determined to lie in the ranges $[0.37,0.59]$ and $[0.03,0.19]$, respectively.

\begin{figure}[tb]
\centering	
\includegraphics[width=0.38\textwidth]{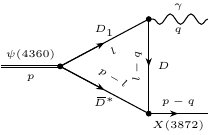}
    \caption{Feynman diagram for $\psi(4360)\to\gamma X(3872)$. The charge-conjugated diagram is not shown.}
    \label{Fig:decay_psi_X3872}
\end{figure}

The decay of $\psi(4360)\to \gamma X(3872)$ can be estimated using the triangle loop diagram in Fig.~\ref{Fig:decay_psi_X3872}. The amplitude is
\begin{align}
{\cal M}_{X}=\sum_{a=u,d}\frac{x_{\nr}g_{\nr}^{\prime}c_a |\bs q|N_3}{2\sqrt 3}\varepsilon_{ijk}\varepsilon_i^*(\psi)\varepsilon_j(\gamma)\varepsilon_k(X)I(\bs q),
\label{Eq:amp_decay_X}
\end{align}
where $N_3=8m_{D_1}m_{\olsi D^*}m_D\sqrt{m_{\psi}m_X}$ is the nonrelativistic normalization factor with the $X(3872)$ mass $m_X$, and $I(\bs q)$ denotes the loop function in Eq.~\eqref{Eq:triangle_loop}. For the effective couplings $g_{\nr}^{\prime}$ of $X(3872)$ to $D^0\bar D^{*0}$ and $D^+D^{*-}$, we take the values 
$g_{\nr, 0}'=(0.26 \pm 0.02) e^{(0.02 \pm 0.01) i}~\mathrm{GeV}^{-1/2}$ and $g_{\nr, \pm}' =(0.16 \pm 0.01) e^{(0.02 \pm 0.01) i}~\mathrm{GeV}^{-1/2}$ determined recently in Ref.~\cite{Ji:2025hjw}. 
We fix the coupling constants for the E1 transition to $c_u=0.48$ and $c_d=0.11$, and then calculate the decay width for $Y(4360)\to \gamma X(3872)$.

\begin{figure}
    \centering	
    \includegraphics[scale=0.8
    ]{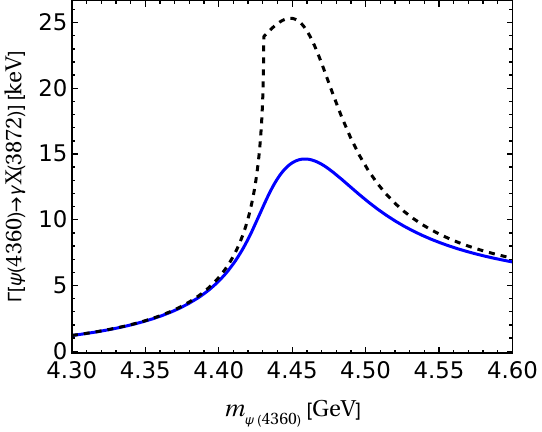}
    \includegraphics[scale=0.8
    ]{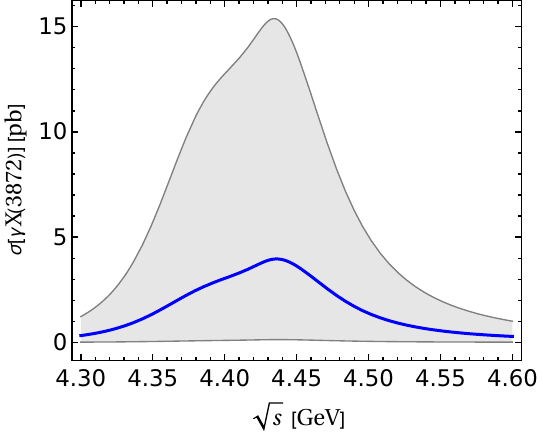}\hfill
    \caption{Radiative decay width of $\psi(4360)\to\gamma X(3872)$ and production rate of $X(3872)$ through the process $e^+e^-\to\psi(4360)\to\gamma X(3872)$. In the left panel, the solid and dashed lines represent the decay width when the decay of $D_1$ is included and excluded, respectively, obtained using central values of all parameters. In the right panel, the gray band indicates the total uncertainty, obtained by adding in quadrature the contributions from the parameters $f_{\psi(4360)}, g_{\nr,0}^{\prime}, g_{\nr,\pm}^{\prime},$ the mass and decay width of $\psi(4360)$, and the higher-order $\mathcal{O}(\gamma/\beta)$ term in Eq.~\eqref{Eq:coupling}.
    \label{Fig:decay_psi_X3872_pred}}
\end{figure}
The result is depicted in the left panel of Fig.~\ref{Fig:decay_psi_X3872_pred}. For the production of $X(3872)$ in the process $e^+e^-\to \psi(4360)\to \gamma X(3872)$, as shown in the right panel of Fig.~\ref{Fig:decay_psi_X3872_pred}, we estimate that the maximum cross section $\sigma^{\text{max}}[\gamma X(3872)]=(4_{-4}^{+11})$~pb at $\sqrt s\approx4436$~MeV.
The contribution from the $\psi(4415)\to\gamma X(3872)$ process is anticipated to be significantly smaller. This is because it would require replacing $D_1$ in Fig.~\ref{Fig:decay_psi_X3872} by $D_2$, and the $D_2\to D\gamma$ transition is expected to be much weaker due to the $D$-wave suppression. Consequently, the contribution from the intermediate $\psi(4415)$ in the production of $\gamma X(3872)$ has been disregarded in the above estimation.

\bibliography{refs.bib}

\end{document}